\documentclass[manuscript]{aastex61}

\usepackage{CJK}

\received{October xx, 2017}
\revised{xxx xx, 2017}
\accepted{xxx xx, 20xx}
\submitjournal{AJ}

\shorttitle{SDSSJ1156-0207}
\shortauthors{Lee}

\begin{document}
\begin{CJK*}{UTF8}{bkai}
\title{SDSSJ1156-0207: A 0.54 + 0.19 M$_\odot$ double-lined M dwarf eclipsing binary system}

\correspondingauthor{Lee, Chien-Hsiu}
\email{leech@naoj.org}

\author{Lee, Chien-Hsiu (李見修)}
\affiliation{Subaru Telescope, National Astronomical Observatory of Japan, 650 North A'ohoku Place, Hilo, HI 96720, USA}



\begin{abstract}
  Eclipsing binaries are instrumental to our understanding of fundamental stellar parameters. With the arrival of
  ultra-wide cameras and large area photometric monitoring programs, numerous eclipsing binaries systems have been
  reported photometrically. However, due to the expensive efforts to follow up them spectroscopically,
  most of their basic properties remain unexplored. In this paper we exploited the eclipsing binary light curves delivered
  by the all-sky catalina sky surveys, in tendem with the single shot spectroscopic survey from SDSS, and identify a double-lined
  M-dwarf eclipsing binary SDSSJ1156-0207. Because this system is very faint (V=15.89 magnitude), we obtained follow-up
  radial velocity measurements using Gemini Multi-Object Spectrograph onboard Gemini north telescope. This provides us a spectral
  resolution R$\sim$4000, enabling us to determine the mass and radius of each stellar components when jointly fitted with light curve.
  Our best-fit results indicate that both components are of M dwarf, with the primary component to be 0.54$\pm$0.20 M$_\odot$ and 0.46$\pm$0.08 R$_\odot$, while the
  secondary component to be 0.19$\pm$0.08 M$_\odot$ and 0.30$\pm$0.08
 R$_\odot$. High resolution spectroscopic observations in the future will help
  pin down the stellar parameters, providing insights to the stellar models at low mass regimes,
  as well as sheding lights on the internal structure of close-in low mass objects and their inflation mechanism.
\end{abstract}




\section{Introduction}

M dwarfs span the cool, low mass end of the main sequence. They are 
the most abundant stars in the Milky Way. However, they are difficult to discover and characterize because of their faintness. Thanks to large sky surveys,
e.g. SDSS \citep{2000AJ....120.1579Y},
several thousands of M dwarfs have been charted \citep{2011AJ....141...97W}.
However, only a few of them are well-studied, where their masses, radii, and temperatures are known to few percent level \citep{2010A&ARv..18...67T}.

Observational results of M dwarfs also cast challenges to the models of low mass stars. It has been shown that theoretical modeling often underestimates the size and 
overestimates the temperature of M dwarfs \citep{2012ApJ...757..112B}.
As shown by \cite{2015MNRAS.451.2263Z}, the discrepancies are prominent among very low mass stars
(VLMS, M$<$0.3M$_\odot$). This is because at such low masses the stars become completely convective and are difficult to model.
Furthermore, these low mass stars are magnetically active, 
which links to their inflation \citep{2013ApJ...779..183F,2014ApJ...789...53F}.

With increasing numbers of exoplanets found to be around low mass stars \citep{2013ApJ...767...95D}, there are growing interests to understand the properties of low mass stars. This is especially the case of planets discovered by the transit method,
because we require accurate stellar 
parameters to estimate exoplanet properties. 
Thus, improving VLMS models, or deriving
relations for mass, radius and effective temperature empirically \citep[e.g.][]{2000A&A...364..217D,2016AJ....152..141B,2015ApJ...804...64M,2015ApJ...800...85N} are highly demanded. 

Eclipsing binaries can deliver direct measurement of the mass,
radius, and effective temperature of stars. From eclipsing binary light curves
we can determine the period, inclination angle, eccentricity, mass
ratio, and fractional radius. With spectroscopic observations, 
we can estimate the mass, radius, and temperature of individual stars.
We note that while there is a degeneracy of extracting both temperature and the gravity
(log$g$) from spectra of a single star, it is not an issue in the case of eclipsing binaries.
This is because we can
determine the mass and radius independently combining the radial velocity information and
light curves. We can therefore
break the degeneracy and estimate the temperature. There have been seminal works to
provide benchmark M-M binary systems, such as CM Draconis (Morales et al., 2009, ApJ, 691, 1400), KOI-126 (Carter et al., 2011, Science, 331, 562), Kepler-16 (Bender et al., 2012, ApJ, 751, L31),  LSPM J1112+7626 (Irwin et al., 2011, ApJ, 742, 123).

Taking advantage of eclipsing binaries composed of VLMS, \cite{2015MNRAS.451.2263Z}
  presented studies from 9 VLMS, including the abovementioned M-M binary systems, with masses between 0.14 to 0.28 M$_\odot$.
  Their results suggest larger radius and lower temperature then those predicted by stellar models.
  To better understand the discrepency between the theoretical modeling and observations, and to
  derive mass vs. radius and mass vs. temperature relations empirically for
  VLMS, we need to determine stellar parameters for a larger sample of VLMS, which is only possible via eclipsing binaries.

Further more, short-period binaries presented in this study tend to be inflated due to tidal locking. Though they are
    not ideal for empirical calibrations, they can provide insights to inflation mechanisms and probe the interior structures of M dwarfs.
  
In this work we present a double-lined, M dwarf
eclipsing binary discovered from cross matching Catalina Sky Surveys and Sloan Digital Sky Survey. The physical properties of this system are further
characterized using Gemini telescope.
This paper is organized as follows: in \textsection \ref{sec.data} we describe the photometric and
spectroscopic data we used. Our analysis is shown in 
\textsection \ref{sec.ana} and discussed in \textsection \ref{sec.dis}, followed by a summary in \textsection \ref{sec.sum}.

\section{Data}
\label{sec.data}
\subsection{Eclipsing binaries from Catalina Sky Survey}
Starting from 2004, the Catalina Sky Survey (CSS) made use of three telescopes to survey the sky with -75 $<$ Dec $<$ 70 deg, with the Catalina Schmidt Telescope (0.7m, using a 8.2 deg$^2$ camera) and the Mount Lemmon Telescope (1.5m, using a 1 deg$^2$ camera) in Arizona, USA, as well the Siding Spring Telescope (0.5m, equipped with 4.2 deg$^2$ FOV) from Australia. 
The original goal of CSS is to discover hazardous near Earth objects (NEOs). To maximize the throughput, the CSS observations were conducted without filters. The exposures were taken in quad manner, with cadence of 10-min among images, as well as an exposure time of 30-sec. The photometry are done using SExtractor \citep{1996A&AS..117..393B}. Due to the high cadence, as well as the wide survey area, CSS data are also pivotal to time-domain studies. This leads to the Catalina Real-time Transient Survey \citep[CRTS][]{2009ApJ...696..870D}.

\begin{figure*}
 \centering
  \includegraphics[scale=1.3]{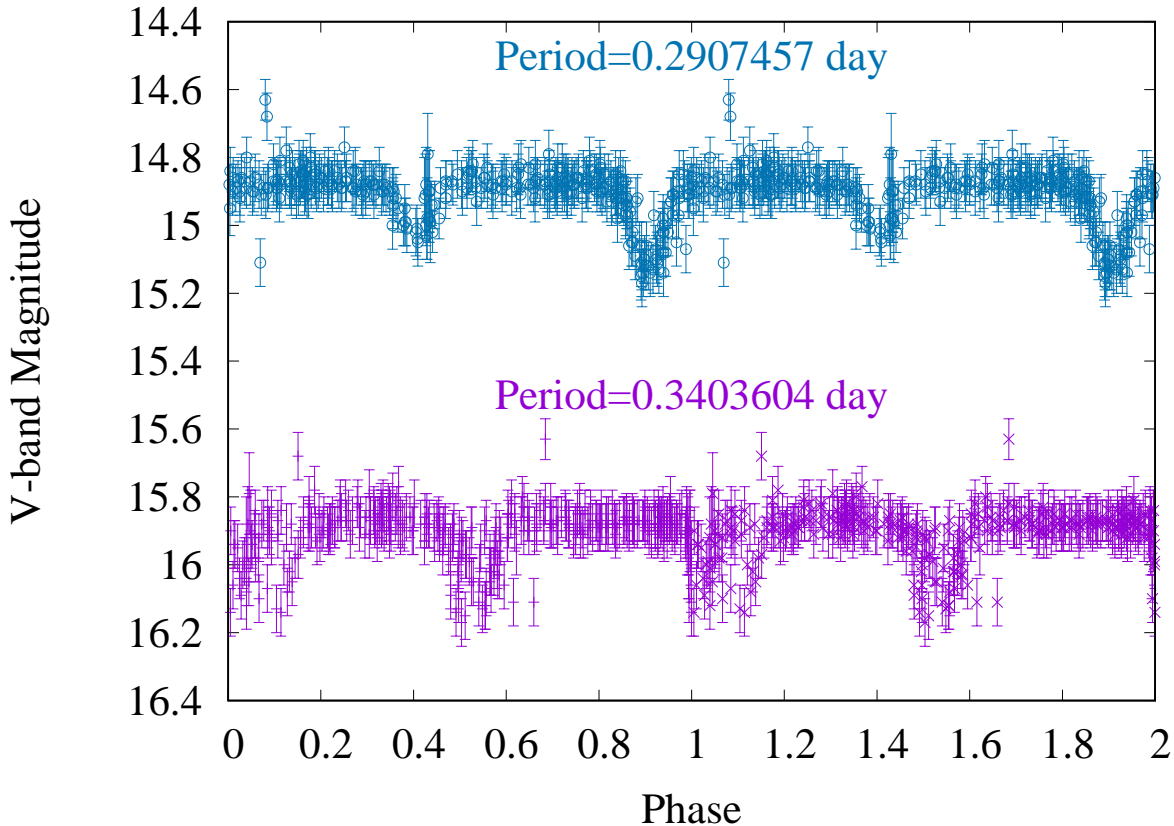}
  \caption{Light curve of SDSSJ1156-0207 folded with different periods. Period estimate from the CSDR2 (0.3403604 days) is shown in violet, while period estaimte from our AoV analysis is shown in blue (0.2907457).}
  \label{fig.flc}
\end{figure*}

The current Catalina Surveys Data Release\footnote{http://nesssi.cacr.caltech.edu/DataRelease/} contains data obtained by the Catalina Schmidt Telescope (with a pixel scale of 1.5''/pixel) obtained in 2005-2013, with the light curves magnitudes converted to V-band.
In the first data release (CSDR1), \citep{2013ApJ...763...32D} presented $\sim$ 200 million sources with 12$<$V$<$20 magnitudes.
\cite{2014ApJS..213....9D} further searched variables in CSDR1 sky regions between -22 $<$ Declination $<$ 65 deg, and
discovered $\sim$ 47,000 periodically varying objects. They visually inspected the
light curves and classified 4683 detached eclipsing binaries. This is largest all-sky eclipsing binary catalog
to-date, providing a reservoir of time-series photometry.
We note that while CSS catalog provides period estimate (0.3403604 days) using Lomb-Scargle periodogram, it is not very accurate (see Fig. \ref{fig.flc}. To determine period precisely, we empolyed the Analysis of Variance method (Schwarzenberg-Czerny, 1989), and estimate a period of 0.2907457 days. To estimate the period error, we use bootstrapping by replacing 1/5 of the data points in the light curve, and repeat the bootstrapping test for 50 times. This yields a period error estimate of 9$\times$10$^{-7}$ days, as presented in Table \ref{tab.phot}.

\subsection{Spectroscopic M dwarfs from SDSS}
As M dwarfs are very faint in the optical wavelength, previous searches were limited to infrared imaging studies,
such as \cite{2011AJ....142..138L,2013PASP..125..809T}.
Thanks to the advent of in wide-field cameras, multi-object spectrographs, and all sky surveys with
medium-size telescopes, tens of thousands of M dwarfs have been confirmed by optical spectra.
For example, \cite{2011AJ....141...97W} took the
advantage of Sloan Digital Sky Survey \citep[hereafter SDSS][]{2000AJ....120.1579Y} to identify and
classify $\sim$ 70,000 M dwarfs. Their sample were drawn from $\sim$ 120,000 M dwarf
candidates pre-classified by
Hammer spectral typing facility \citep{2007AJ....134.2398C}, further excluding low S/N spectra and
removing contaminations from extra-galactic objects, K/L dwarfs, and white dwarf - M dwarf pairs. In the
end they presented 70,841 M dwarfs with spectral type error within 1 subtype.
This is the largest and purest spectroscopic M dwarf catalog to-date,
providing us a firm basis to cross-matching with eclipsing binaries charted by other time-domain surveys.

\begin{figure*}
 \centering
  \includegraphics[scale=0.7]{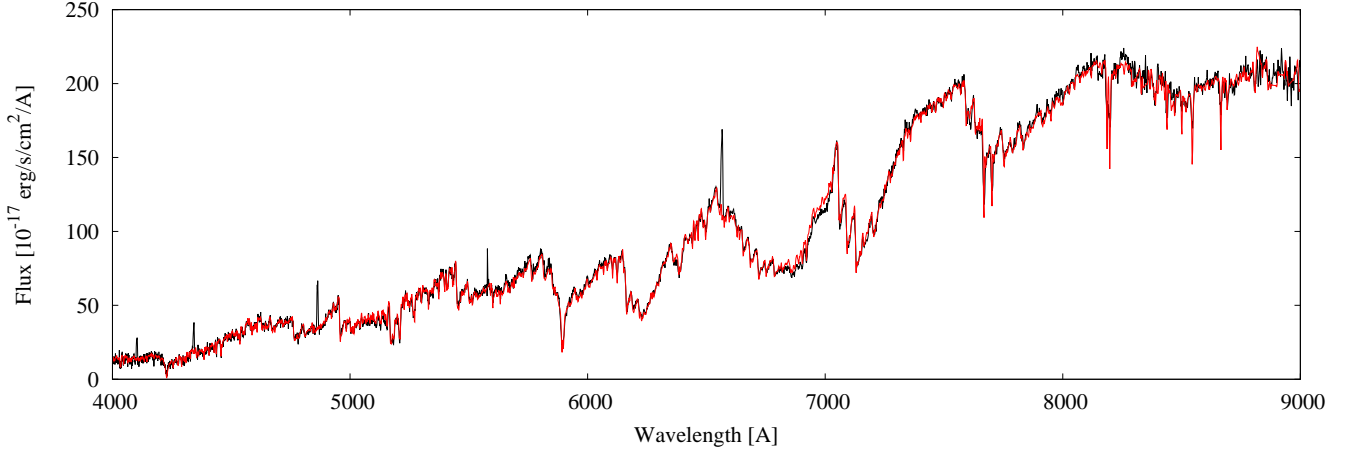}
  \caption{SDSS spectrum of SDSSJ1156-0207 (shown in black) overlaid with the best-fit spectrum of a M3 dwarf template (shown in red). We note that the SDSS fiber has a radius of 3'', hence the SDSS spectrum is partially contaminated by nearby star as we shown in section \ref{sec.dis}.}
  \label{fig.sdss}
\end{figure*}

In order to identify candidate VLMS systems in eclipsing binaries, we matched the M dwarfs in \cite{2011AJ....141...97W} to the CSS eclipsing binary catalogue. We only looked for M dwarfs with spectral classification later than M3, which have stellar masses$<$0.3M$_\odot$ \citep[][]{1996ApJ...461L..51B}. In the end we identified
two VLMS M dwarfs in eclipsing systems: CSS114804.3+244132/SDSSJ114804.35+255132.6 in Lee (2017), as well as SDSSJ1156-0207 as presented here.
The photometric and astrometric parameters, as well as
the SDSS spectrum of SDSSJ1156-0207, and shown in Table \ref{tab.phot} and
in Fig. \ref{fig.sdss}.  

\begin{table}
\centering
\caption{Photometric and astrometric parameters of SDSSJ1156-0207}
\begin{tabular}[t]{ll}
  \hline
  \hline
  Parameter & Value \\
  \hline
  RA (J2000) & 11:56:41.68 \\
  Dec (J2000) & -02:07:03.8 \\
  $\mu_{RA}$ [mas/yr] & 9.0 \\
  $\mu_{Dec}$ [mas/yr] & -34.0 \\
  u [mag] & 20.353$\pm$0.049 \\
  g [mag] & 17.874$\pm$0.006 \\
  r [mag] & 16.381$\pm$0.004 \\
  i [mag] & 15.173$\pm$0.004 \\
  z [mag] & 14.503$\pm$0.005 \\
  J [mag] & 13.137$\pm$0.046 \\
  H [mag] & 12.572$\pm$0.044 \\
  K [mag] & 12.262$\pm$0.025 \\
Period [day] & 0.2907457 $\pm$ 9$\times$10$^{-7}$ \\
  \hline
\hline
\end{tabular}
\label{tab.phot}
\end{table}

\subsection{GMOS follow-up spectra}
\label{sec.gmos}
SDSSJ1156-0207 is rather faint (V$\sim$16 mag); it also exhibits a very short period (P$\sim$0.3 days).
In this regard, we need to reach good S/N $\sim$50-100) within less than an
hour (0.1 periods). Further more, we also need spectral resolution at least several $\AA$ per pixel to
estimate the mass and radius. In this regard, only 8-m telescope class telescopes can deliver the desired spectral resolution and signal-to-noise ratio in the short time window. We carried out spectroscopic follow-up with GMOS \citep{2004PASP..116..425H} mounted on the Gemini telescope (Program ID GN-2016-FT16). We
used the R831 grating along with a 0.5''-slit, reaching a resolution R=4396.
The central wavelength of the spectrum is set to be 7000$\AA$, with a coverage of $\sim$2000$\AA$.

The observations were carried out on 2016 April 2nd and 14th during the expected radial velocity maxima around light curve phase 0.25 and 0.75. In principle we only need one observations to measurement the radial velocity maximum, nevertheless we obtain two exposures at each maximum, with the second exposure serves as a sanity check, ensuring we
obtain consistent radial velocity measurement at the same maximum. Each exposure has 10 minutes integration to reach sufficient S/N. Data reduction were carried out using
dedicated IRAF\footnote{http://iraf.noao.edu} GMOS package\footnote{http://www.gemini.edu/sciops/data-and-results/processing-software} (v1.13) in a standard manner. Each spectrum was bias subtracted, flat fielded, sky subtracted,
and wavelength calibrated using CuAr lamp. 

To extract radial velocity information, we make use of the H$\alpha$ emission line at $\lambda\lambda$ 6562.8$\AA$.
The H$\alpha$ emissions from both stellar component are clearly resolved. We then
fit the H$\alpha$ line profile using two Gaussian functions. From the Gaussian fit, we
obtain the radial velocity of each stellar component, as shown in Table \ref{tab.rv}, with estimated error of 15 km/s based on the
spectral resolution delivered by the instrument (i.e. 3.4$\AA$ per pixel).


\begin{table}
\centering
\caption{Radial velocity from GMOS observations.}
\begin{tabular}[t]{lrr}
  \hline
  \hline
Epoch & RV$_1$ & RV$_2$ \\ 
(HJD) & [km/s] & [km/s] \\ 
\hline
2457480.335324 &  63.99$\pm$6.7 & -165.02$\pm$10.7 \\
2457480.343681 &  63.08$\pm$6.7 & -149.93$\pm$7.4 \\
2457592.427153 & -34.74$\pm$1.1 &  134.39$\pm$10.9 \\
2457592.435509 & -37.48$\pm$3.4 &  121.59$\pm$2.2 \\
\hline
\hline
\end{tabular}
\label{tab.rv}
\end{table}

\begin{figure*}
  \centering
  \includegraphics[scale=0.35]{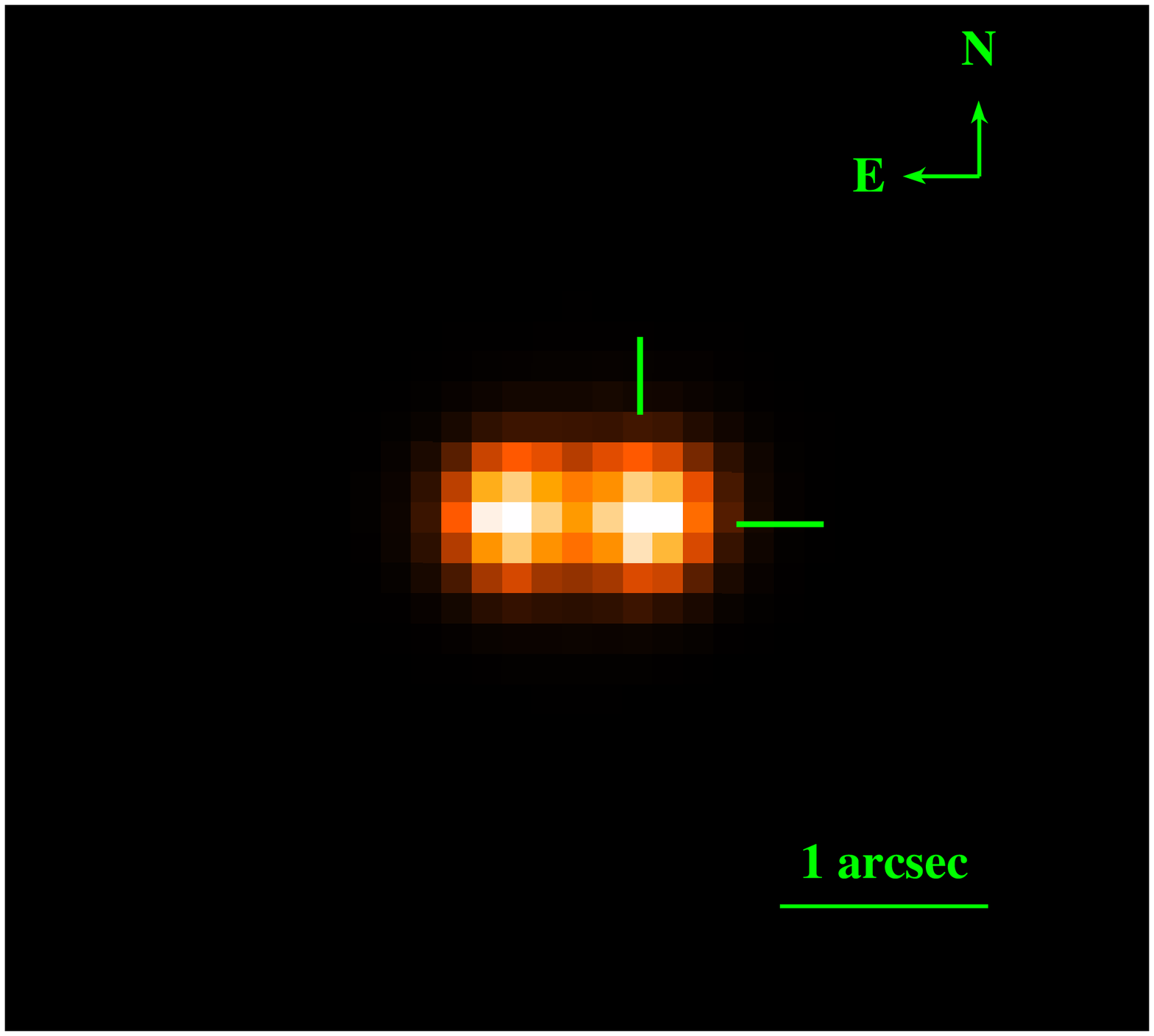}
  \includegraphics[scale=0.4]{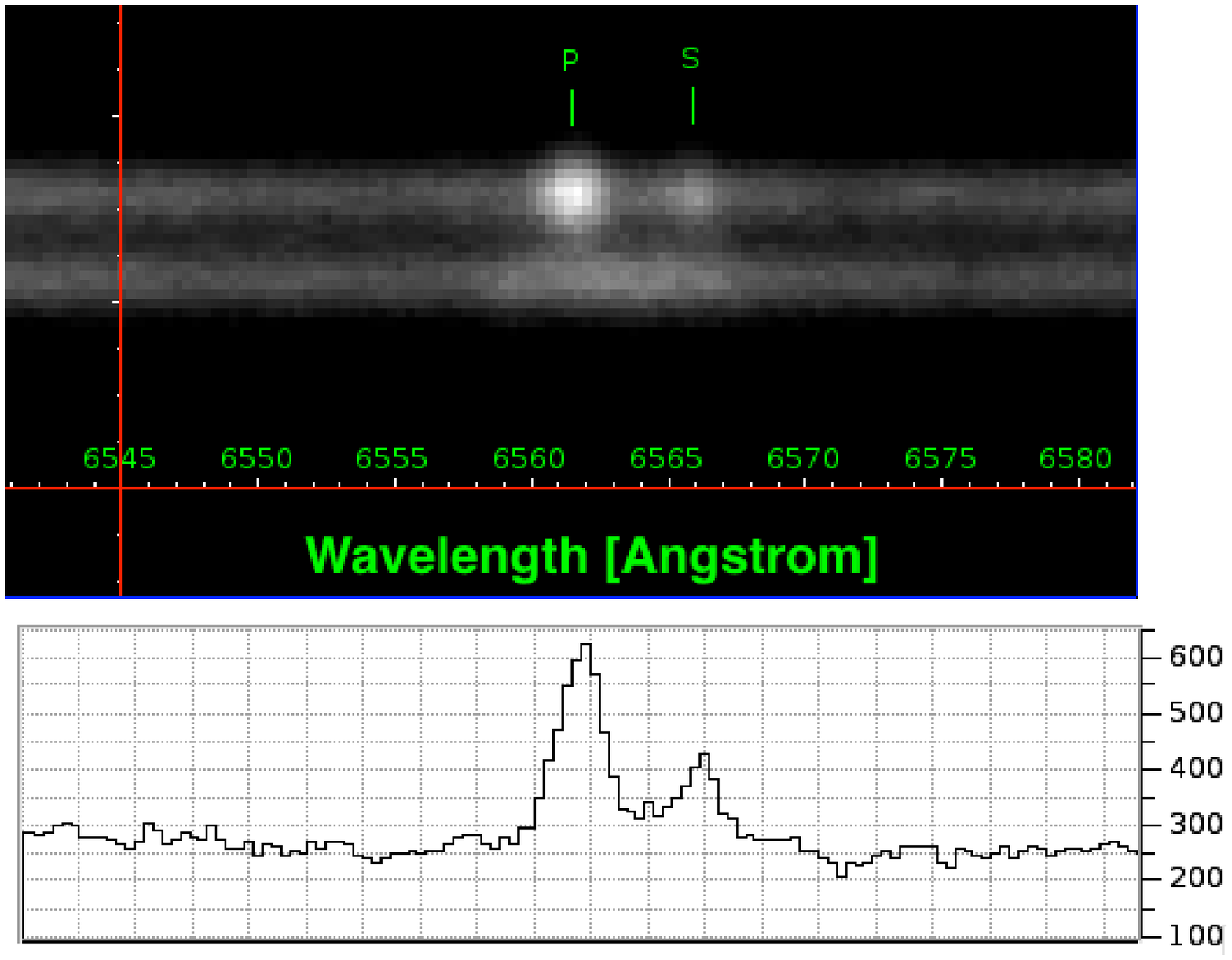}
  \caption{Left: GMOS acquisition image of the binary system. The binary system is indicated
    by the green mark, where the third object (at a separation of 0.5 arcseconds) is well resolved. The Gemini spectra were carried out with a position
    angle of 90 degrees east of north, hence were resolve both the eclisping binary system and the third
    object into spectra. Right: GMOS spectra. The top panel shows the eclipsing binary spectrum (upper) and the third light object spectrum (lower). The primay (P) and secondary (S) component of the eclipsing binary system are indicated by the green label. The bottom panel shows the relative flux (in ADU) of the
  H$\alpha$ emission line from the primary and secondary component of the ecliping binary system.}
  \label{fig.vlms2}
\end{figure*}

We also exploit the acquisition imaging of the Gemini spectroscopic observation
(see Fig. \ref{fig.vlms2}, and found a third object $\sim$0.5 arcseconds
away from the binary system. Such small separation is not resolvable from the poor spatial resolution of CSS. We also checked the Gemini follow-up spectra. As our spectra were obtained with position angle of 90 degrees, east of north, we were able to resolve both the eclipsing binary system and the third light object into spectra, as shown in Fig. \ref{fig.vlms2}. The radial velocity presented here were measured using the
spectra of eclipsing binary system alone, without contaminations from the third light object.

\subsection{WIYN/ODI high precision light curve}
\label{sec.wiyn}
Whilte the discovery light curve from CSS is sufficient to identify the eclipsing binary nature, it
is not ideal for light curve analysis especially due to the small aperture size and the large uncertainty
in the photometry. In this regard, we thus obtained further high precision time-series photometry of
SDSSJ1156-0207 using the One Degree Imager (ODI) on the WIYN telescope. The observations were carried out
on February 3rd 2017 using i-band filter, with an individual exposure time of 120-sec. The median seeing
was between 0.7'' and 0.8''.

The raw data were then de-trended, astrometric and photometric calibrated
using the ODI Pipeline, Portal, and Archive (ODI-PPA) system\footnote{The ODI Pipeline, Portal, and Archive (ODI-PPA) system is a joint development project of the WIYN Consortium, Inc., in partnership with Indiana University's Pervasive Technology Institute (PTI), University of Wisconsin, and with the National Optical Astronomy Observatory Science Data Management (NOAO SDM) Program.}. We retrieved the calibrated data via ODI-PPA, and performed
aperture photometry using the SExtractor package (Bertin \& Arnouts 1996). The light curve from WIYN/ODI is
further discussed in section \ref{sec.ana}.

\section{Analysis}
\label{sec.ana}
\subsection{Dynamical modeling}

To start, we analyse the CSS light curve of SDSSJ1156-0207 using
the Detached Eclipsing Binary Light curve fitter \citep[DEBiL,][]{2005ApJ...628..411D}.
DEBiL fits a binary light curve assuming a simple geometry: it guesses the initial values of
the parameters
using approximated analytic formula \citep[see e.g.][]{2003ApJ...585.1038S},
and fits the following parameters:\\
1. The fractional radius ($\frac{R_1}{a}$,$\frac{R_2}{a}$), expressed in terms of the
semi-major axis $a$.\\
2. The magnitudes of both stellar components (mag$_1$,mag$_2$).\\
3. The orbital eccentricity ($e$), inclination angle ($i$), and argument of periastron ($\omega$).\\

DEBiL performs light curve fitting iteratively, minimising the $\chi^2$ using the downhill simplex method,
and finds the $\chi^2$ minima \citep{1965CJ...7...308N} using simulated annealing \citep{1992nrfa.book.....P}.
We then pass the best-fit DEBiL parameters to JKTEBOP \citep{2004MNRAS.351.1277S} for further dynamical analysis.

\begin{figure*}
 \centering
  \includegraphics[scale=1.2]{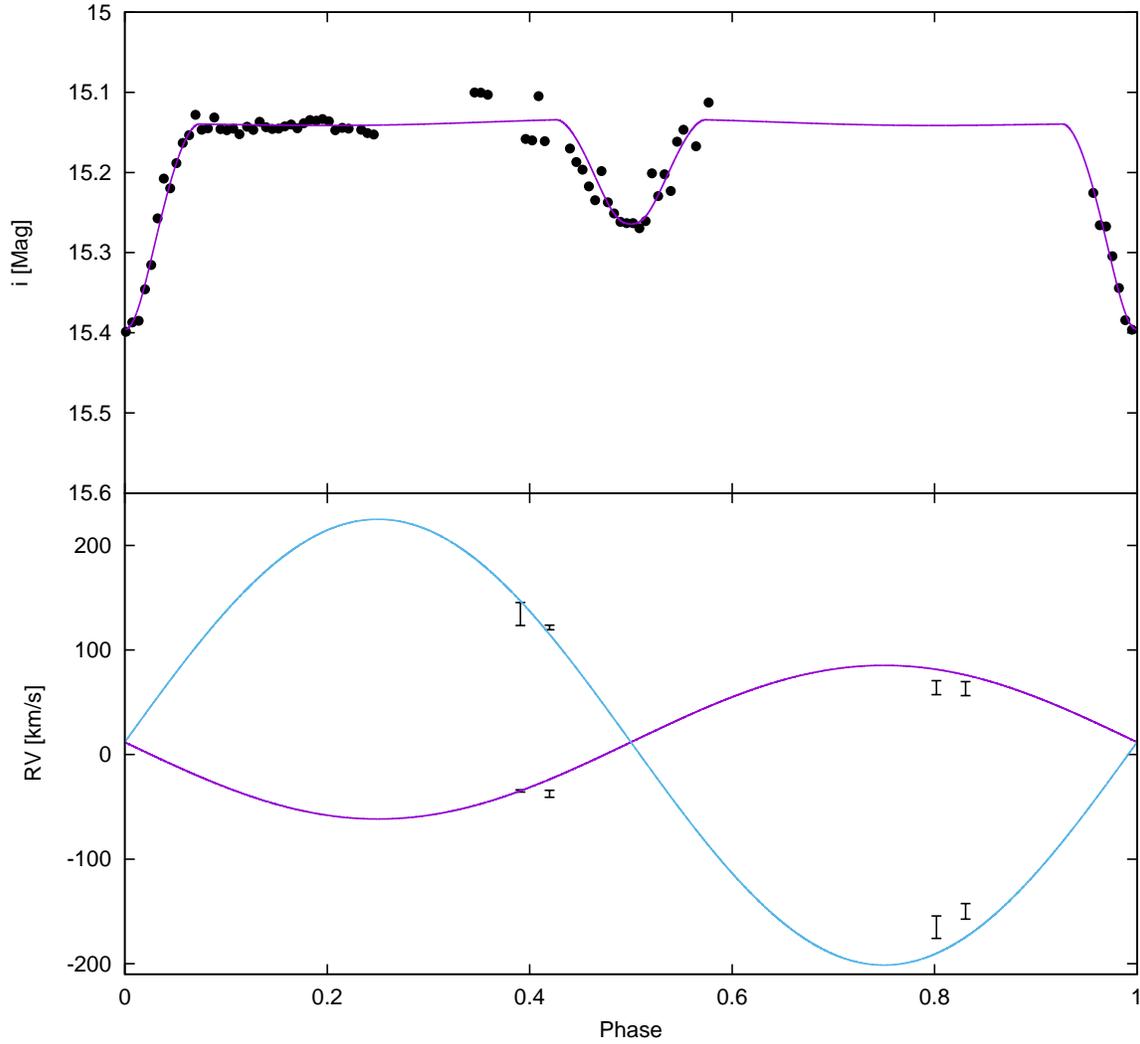}
  \caption{Joint light curve and radial velocity analysis using JKTEBOP \citep{2004MNRAS.351.1277S}. Top pannel shows the WIYN/ODI light curve (black points) over-plotted with the best-fit model (red line). Bottom panel shows the radial velocity measurements from GMOS (black points), with the best-fit model shown in blue (primary component) and red (secondary component) lines.}
  \label{fig.fit}
\end{figure*}

JKTEBOP \citep{2004MNRAS.351.1277S} uses the EBOP code from \cite{1981AJ.....86..102P}. It is capable of
fitting the light curves and 
radial velocity measurements at the same time, to determine the mass and radius of each stellar component.
The following parameters are used during the fitting process: primary eclipse timing (t$_0$), radius ratio (R$_2$/R$_1$), radius sum in terms of semi-major axis ((R$_1$+R$_2$)/$a$), light ratio(L$_2$/L$_1$), inclination angle ($i$), as well as semi-amplitudes of the radial velocity curve (K$_1$ and K$_2$)and the systemic velocity $\gamma$. For dynamical analysis purpose, we only use the high precision
  i-band light curve from WIYN/ODI. In addition, in the WIYN/ODI image we can partially resolve the nearby star contaminating the photometry
  of SDSSJ1156-0207 (see Fig. \ref{fig.vlms2}). This allows us to we put a strong constraint on the 3rd light ratio. We thus fixed
  the 3rd light ratio to be 0.42 for the JKTEBOP analysis. We note that the period is very short, indicating that the binary is tidally locked and the orbital is circularized. We thus fix the eccentricity to be zero and only fit a non-eccentric orbit. During the fitting, we assume a limb darkening in quadratic form, with a=0.37 and b=0.49 for both stellar components. This value is taken from Claret (2004), assuming log\textit{g}=5.0 and T$_{eff}$=3500 for SDSS i-band filter. We also assume a gravity darkening =0.0 and reflection =0.0 for both stellar components.

We show the best-fit JKTEBOP results in Fig. \ref{fig.fit} and
Table \ref{tab.fit}.
We note that there is a general degeneracy of determine the radii in non-eccentric eclipsing binary systems. While the sum of radii can be well determined from the light curve fit, the ratio of radii
is much less uniquely determined, see e.g. Irwin et al. (2011) and references therein. Hence the radius ratio presented here must be regarded with caution.

\begin{table}
\centering
\caption{Best-fit parameters for SDSSJ1156-0207}
\begin{tabular}[t]{lr}
  \hline
  \hline
  Parameter & Result \\
  \hline
  \multicolumn{2}{c}{\underline{\it Modeled parameters}} \\
 
t$_0$ [HJD] & 2457788.2927$\pm$0.0002 \\
L$_2$/L$_1$ & 0.54$\pm$0.03 \\
(R$_1$+R$_2$)/$a$ & 0.46$\pm$0.02\\
R$_2$/R$_1$ & 0.64$\pm$0.07\\
$i$ [deg] & 83.5$\pm$2.3\\
K$_1$ [km/s] & 73.4$\pm$2.7\\
K$_2$ [km/s] & 213.1$\pm$4.3\\
$\gamma$ [km/s] & 11.8$\pm$1.6 \\
\hline
\hline
\multicolumn{2}{c}{\underline{\it Derived parameters}} \\
\hline
M$_1$ [M$_\odot$] & 0.54$\pm$0.20\\
M$_2$ [M$_\odot$] & 0.19$\pm$0.08\\
R$_1$ [R$_\odot$] & 0.46$\pm$0.08\\
R$_2$ [R$_\odot$] & 0.30$\pm$0.08\\
a [AU] & 0.0077$\pm$0.0008\\
\hline
\hline
\end{tabular}
\label{tab.fit}
\end{table}

\subsection{SED fitting}
\begin{figure*}
 \centering
  \includegraphics[scale=1.2]{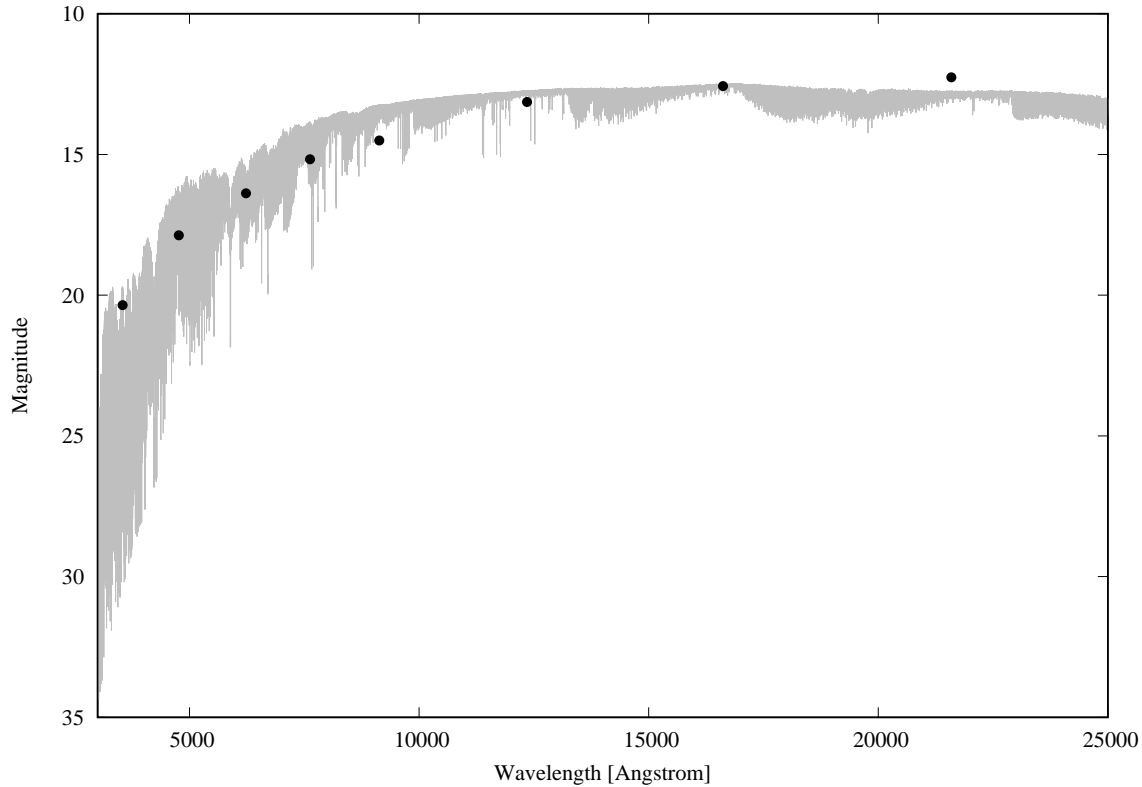}
  \caption{Fitting the broad-band photometry with BT-Settl synthetic spectrum. The best-fit BT-Settl spectrum (with T$_{eff}$ = 3,000 K) is shown in gray. The broad-band photometry from SDSS ugriz and 2MASS JHK filters are shown in black points.}
  \label{fig.sed2}
\end{figure*}

To derive the temperature of this eclipsing binary system, We also perform a basic analysis based on the broad-band spectral energy distribution. We fit the SDSS ugriz and 2MASS JHK broad-band magnitudes using a grid of M dwarf
  model spectra. These model spectra are drawn from the BT-Settl model (Allard et al. 2012), with log$g$=5, solar
  metallicity, and a range of effective temperature from 2500 to 4000 K, with a grid size of 100 K. Our best-fit BT-Settl (as shown in Fig. \ref{fig.sed2}
suggests the system has an effective temperature of 3000 K. To further derive the effective temperature of the primary and secondary stars, we assume the two stars are blackbodies, i.e. $L\propto R^2T^4$. With the light ratio and the radius ratio in hand, we estimate an effective temperature of 3101 K for the primary and 2899 K for the secondary, respectively.

\section{Discussion}
\label{sec.dis}

\begin{figure*}
 \centering
  \includegraphics[scale=1.2]{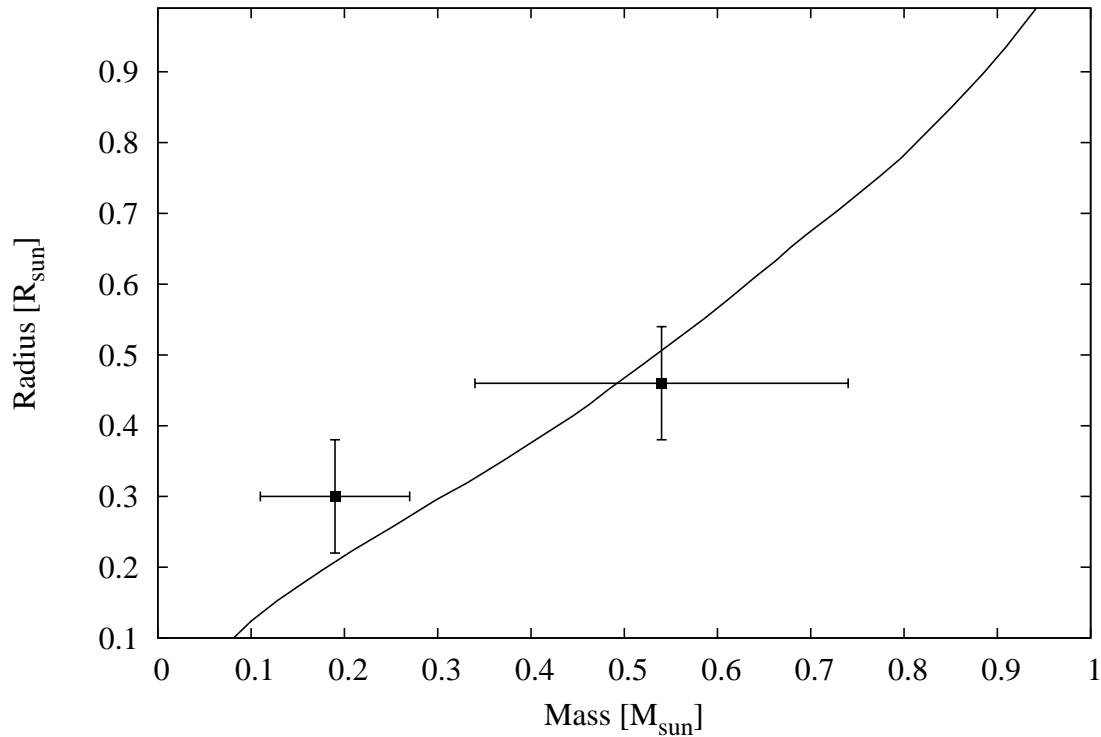}
  \caption{Mass-radius relation of M dwarfs (Baraffe et al. 1998). Both stellar components of SDSSJ1156-0207 are marked in black squares with error bars. The primary star is consistent with the theoretical mass-radius relation, whilte the secondary star appears to be inflated.}
  \label{fig.mr}
\end{figure*}

The best-fit model from JKTEBOP indicates M$_1$=0.54 M$_\odot$, M$_2$=0.19 M$_\odot$, R$_1$=0.46 R$_\odot$, 
and R$_2$=0.30 R$_\odot$.
We compare our results with stellar models of low mass stars by Baraffe et al. (1998), as shown in Fig. \ref{fig.mr}.
We note that our secondary (M$_2$ = 0.19 M$_\odot$) is among the lowest mass of known M-M dwarf binaries. The secondary star appears to be inflated, but it is still consistent with the theoretical predictions considering the large uncertainty in our radius measurements. Future spectroscopic follow-up, especially high resolution spectrograph, will help narrow down the estimated uncertainties.

\section{Summary}
\label{sec.sum}
We perform preliminary analysis of the low mass eclipsing binary
SDSSJ1156-0207, found by cross-matching CSS eclipsing binaries and SDSS spectroscopic M dwarf.
We further obtained spectra with GMOS mounted on the Gemini North telescope,
detecting the emission lines from each of the stellar components. The spectroscopic observations also
provide the radial velocity maxima.
We analyzed the photometric and spectroscopic measurements simultaneously using JKTEBOP. Our results
suggest the secondary component is inflated, while the primary has a radius consistent with the empirical mass-radius relation. The secondary's inflation might be due to tidal locking, which enhances stellar activity and inhibites convection. This is not unusual in short period binaries. High resolution spectroscopy in the future will help narrow down the basic properties of this system. Further H$\alpha$ observations will shed light on the stellar activity, providing constraints on the inflation mechanism due to tidal-locking.

\acknowledgments

We are indebeted to the referee, whose insightful comments great improved the manuscript.

Data presented herein were obtained at the WIYN Observatory from telescope time allocated to NN-EXPLORE through the scientific partnership of the National Aeronautics and Space Administration, the National Science Foundation, and the National Optical Astronomy Observatory.

The CSS survey is funded by the National Aeronautics and Space Administration under Grant No.
NNG05GF22G issued through the Science Mission Directorate Near-Earth Objects Observations Program.
The CRTS survey is supported by the U.S.-National Science Foundation under grants AST-0909182.

Funding for the Sloan Digital Sky Survey IV has been provided by
the Alfred P. Sloan Foundation, the U.S. Department of Energy Office of
Science, and the Participating Institutions. SDSS-IV acknowledges
support and resources from the Center for High-Performance Computing at
the University of Utah. The SDSS web site is www.sdss.org.

SDSS-IV is managed by the Astrophysical Research Consortium for the 
Participating Institutions of the SDSS Collaboration including the 
Brazilian Participation Group, the Carnegie Institution for Science, 
Carnegie Mellon University, the Chilean Participation Group, the French Participation Group, Harvard-Smithsonian Center for Astrophysics, 
Instituto de Astrof\'isica de Canarias, The Johns Hopkins University, 
Kavli Institute for the Physics and Mathematics of the Universe (IPMU) / 
University of Tokyo, Lawrence Berkeley National Laboratory, 
Leibniz Institut f\"ur Astrophysik Potsdam (AIP),  
Max-Planck-Institut f\"ur Astronomie (MPIA Heidelberg), 
Max-Planck-Institut f\"ur Astrophysik (MPA Garching), 
Max-Planck-Institut f\"ur Extraterrestrische Physik (MPE), 
National Astronomical Observatory of China, New Mexico State University, 
New York University, University of Notre Dame, 
Observat\'ario Nacional / MCTI, The Ohio State University, 
Pennsylvania State University, Shanghai Astronomical Observatory, 
United Kingdom Participation Group,
Universidad Nacional Aut\'onoma de M\'exico, University of Arizona, 
University of Colorado Boulder, University of Oxford, University of Portsmouth, 
University of Utah, University of Virginia, University of Washington, University of Wisconsin, 
Vanderbilt University, and Yale University.

Based on observations obtained at the Gemini Observatory and processed using the Gemini IRAF package,
which is operated by the Association of Universities for Research in Astronomy, Inc., under a cooperative
agreement with the NSF on behalf of the Gemini partnership: the National Science Foundation (United
States), the National Research Council (Canada), CONICYT (Chile), Ministerio de Ciencia, Tecnologiae Innovacion Productiva (Argentina), and Ministerio da Ciencia, Tecnologia e Inovacao (Brazil).

The authors wish to recognize and acknowledge the very significant cultural role and reverence that the
summit of Maunakea has always had within the indigenous Hawaiian community. We are most fortunate to
have the opportunity to conduct observations from this mountain.


\end{CJK*}

\end{document}